\documentclass[twocolumn,prl]{revtex4}
\usepackage{amssymb}
\usepackage[all]{xy}
\usepackage{amsmath}
\usepackage{graphicx}
\newcommand{\be}{\begin{equation}}
\newcommand{\ee}{\end{equation}}
\newcommand{\bea}{\begin{eqnarray}}
\newcommand{\eea}{\end{eqnarray}}

\newcommand{\ra}{\rangle}

\newcommand{\la}{\langle}

\begin{document}


\title{Wave-particle Interactions in a Resonant System of Photons and Ion-solvated Water}

\author{Eiji Konishi\footnote{konishi.eiji.27c@st.kyoto-u.ac.jp}}
\affiliation{Graduate School of Human and Environment Studies, Kyoto University, Kyoto 606-8501, Japan}


\date{\today}

\begin{abstract}
We investigate a laser model for a resonant system of photons and ion cluster-solvated rotating water molecules in which ions in the cluster are identical and have very low, non-relativistic velocities and direction of motion parallel to a static electric field induced in a single direction.
This model combines Dicke superradiation with wave-particle interaction.
As the result, we find that the equations of motion of the system are expressed in terms of a conventional free electron laser system.
This result leads to a mechanism for dynamical coherence, induced by collective instability in the wave-particle interaction.
\end{abstract}

\keywords{Wave-particle interaction; Dicke superradiation; Free electron laser}

\maketitle

The mechanisms behind {\it coherence} in physical systems are mainly classified into two types.
In the first type, coherence is due to long-range order; in the second type, it is due to the collective instability in a many-body system with a long-range interaction.
In laser physics, Dicke superradiation belongs to the first type and the free electron laser (FEL) belongs to the second type\cite{Dicke,Super,FEL,Bonifacio,BookFEL}.
In this Letter, we present a new laser model that incorporates both types of dynamical coherence mechanism from the aspect of wave-particle interactions\cite{Book,FELPR,FEL1,FEL2,FEL3}.
The ingredients of this system are {\it photons}, {\it solvated ions} and {\it water molecules}.

In Ref.\cite{GPV} the many-body system of electric dipoles of water molecules that interact with the electromagnetic field radiated from rotating water molecules was studied using a quantum field theoretical approach based on an analogy to an FEL\cite{FEL,Bonifacio,FELPR,FEL1,FEL2,FEL3,Nature}.
In simplified settings, it was shown that, around an impurity that carries a sizeable electric dipole and induces a static electric field oriented in the $z$-direction, a permanent electric polarization of water molecules in the $z$-direction emerges in the limit cycle of the system, due to the coherent and collective interaction of the water molecules with the selected modes of the radiation field.

Using this result, in this Letter we consider an FEL-like model for a cluster of solvated identical ions with very low, non-relativistic velocities and a uniform direction of motion.
The electric dipoles of the water molecules which solvate each moving ion behave as $XY$ spins which interact directly only with the transverse electromagnetic field radiated from the rotating water molecules.
On the particle side of the wave-particle (i.e., radiation field-particle) interactions, there are two kinds of elements: water molecules and ions around which the water molecules are trapped by a Lennard-Jones potential.

In the FEL model\cite{FEL,Bonifacio,BookFEL,FELPR}, each relativistic unbound electron has the phase degree of freedom of its $XY$ spin-like direction on the $x$-$y$ plane orthogonal to its direction of motion (called the $z$-direction).
Due to the presence of an undulator, that is, a transverse magnetic field created by a periodic arrangement of magnets with alternating poles, the unbound electrons are accelerated, producing synchrotron radiation.

In our case, the variable corresponding to the phase coordinate of an unbound electron is the phase $\theta$ of the direction of the electric dipole vector $\vec{d}$ on the $x$-$y$ plane for each water molecule.
Classically, we have 
\begin{equation}
d_{x}+id_y=d_0|\sin \varphi| e^{i\theta}\;,
\end{equation}
where $\varphi$ is the zenith angle and $d_0=2ed_e$ with $d_e\approx 0.2$ [\AA]\cite{Franks}; however, we consider it quantum mechanically.
Here, we assume that ions move along the $z$-axis with velocity $v\ll c$\footnote{In this Letter, quantities of the order of $(v/c)^n$ ($n\ge 1$) are ignored.} and a static electric field $\vec{E}_0$ is induced in the $z$-direction.
Then, we can invoke the result of Ref.\cite{GPV}, that is, the emergence of a permanent electric polarization of the solvent and non-screening water\footnote{The solvent water obeys an electrostatic ordering mechanism (see the explanation below Eq.(\ref{eq:Hw}))\cite{Water}.} in the $z$-direction.

This Letter is based on the assertion that, {\it{to a good approximation in our system, the radiation field exchanges energy with water molecules only through the excitation and de-excitation of water molecules between the two lowest levels of the internal rotation of the hydrogen atoms of each water molecule around its electric dipole axis.
The energy difference between these two levels is ${\cal E}$, such that ${\cal E}/(\hbar c) \approx 160$ [cm$^{-1}$]}}\cite{GPV,Franks,JPY}.
To describe this {\it resonant} interaction, in the quantum mechanical regime, we introduce the {\it energy} spin variables of the $j$-th water molecule as \footnote{In this Letter, a hat indicates that a variable is a quantum mechanical operator.}
\begin{eqnarray}
\widehat{s}^1_j&=&\frac{1}{2}[|e\ra\la g|+|g\ra\la e|]_j\;,\\
\widehat{s}^2_j&=&\frac{1}{2i}[|e\ra\la g|-|g\ra\la e|]_j\;,\\
\widehat{s}^3_j&=&\frac{1}{2}[|e\ra\la e|-|g\ra\la g|]_j\;,\label{eq:Spin}
\end{eqnarray}
where $|g\ra_j$ and $|e\ra_j$ are, respectively, the low-lying energy ground state and the excitation energy state of the $j$-th water molecule in the two-level approximation, and the superscripts of the energy spins represent fictitious dimensions\cite{PRSR,JPY}.
We introduce an electric dipole moment operator $\vec{\widehat{d}}_j$ for the $j$-th water molecule such that its third axis is the quantization axis of the rotating water molecule.
One of the off-diagonal matrix elements $(\la e|\vec{\widehat{d}}|g\ra)_j=\overline{(\la g|\vec{\widehat{d}}|e\ra)_j}$ of the electric dipole moment operator $\vec{\widehat{d}}_j$ of the $j$-th water molecule is given by\footnote{In its definition, the frame $(\vec{e}_1,\vec{e}_2,\vec{e}_3)$ takes paths in $O(3)$ so that its distribution contains  $(\vec{e}_x,\vec{e}_y,\vec{e}_z)$, $(\vec{e}_z,\vec{e}_x,\vec{e}_y)$, $(\vec{e}_y,\vec{e}_z,\vec{e}_x)$, $(-\vec{e}_x,-\vec{e}_y,-\vec{e}_z)$, $(-\vec{e}_z,-\vec{e}_x,-\vec{e}_y)$, and $(-\vec{e}_y,-\vec{e}_z,-\vec{e}_x)$ and is statistically symmetric with respect to $1$, $2$, and $3$ in the cyclic order.}
\begin{widetext}
\begin{eqnarray}
d_0\int_{-1}^1d(\cos \varphi)\int_0^{2\pi}d\theta (\bar{Y}_{1,1}(\varphi,\theta)(\vec{e}_1\sin\varphi \cos \theta+\vec{e}_2\sin \varphi \sin \theta+\vec{e}_3\cos \varphi)_jY_{0,0}(\varphi,\theta))={d_0}\sqrt{\frac{1}{6}}(-\vec{e}_1+i\vec{e}_2)_j\;.
\end{eqnarray}
\end{widetext}
Then, the electric dipole moment operator can be represented as an off-diagonal matrix in the two-dimensional energy state space:\cite{Dover}
\begin{equation}
\vec{\widehat{d}}_j=\frac{\tilde{d}_0}{2}(-\vec{e}_1(2\widehat{s}^1)-\vec{e}_2(2\widehat{s}^2))_j\;,
\end{equation}
where we define
\begin{equation}
\tilde{d}_0={d_0}\sqrt{\frac{2}{3}} \approx 0.82\cdot d_0\;.
\end{equation}

The Hamiltonian for the interaction between ion-solvated water and the transversal radiation vector field $\vec{\widehat{A}}$ (i.e., the electromagnetic field in the radiation gauge) can be written using the total relevant electric charge current $\vec{\widehat{j}}$ of the solvated ions and water molecules as
\begin{equation}
\widehat{H}_{{\rm int}}^{(p-i-w)}=-\sum_{I=1}^{N}\sum_{i=0}^{n_I}\vec{\widehat{A}}\cdot {\vec{\widehat{j}}}_{{I,i}}\;,\label{eq:Hint}
\end{equation}
where the natural number $n_I$ is the number of water molecules solvating the $I$-th ion and is observed to be $20$ to $40$\cite{Number0,Number1,Number2,Number3}.
The total relevant electric polarization current $\vec{\widehat{j}}$ is the sum of the ions' electric charge currents $\vec{\widehat{j}}_I$ and the $1$,$2$-components of the water molecules' electric polarization currents $\dot{\vec{\widehat{d}}}_{I,i}$:
\begin{eqnarray}
\vec{\widehat{j}}_{I,0}&=&\vec{\widehat{j}}_I\;,\\
\vec{\widehat{j}}_{I,i}&=&(\vec{e}_1 \dot{\widehat{d}}_1+\vec{e}_2 \dot{\widehat{d}}_2)_{I,i}\ \ (i\neq 0)\;,\label{eq:dd}
\end{eqnarray}
where the time derivative $\dot{\vec{\widehat{d}}}_{I,i}$ is taken in the interaction picture.
In the interaction Hamiltonian Eq.(\ref{eq:Hint}), the odd-parity operator $\vec{\widehat{j}}_{I,i}$ ($i\neq 0$) has only off-diagonal matrix elements in the representation where the water molecule's free Hamiltonian in the two-level approximation ${\cal E}\widehat{s}_{I,i}^3$ is diagonal\cite{JPY}.

It is a crucial point that the photon-water molecule part of Eq.(\ref{eq:Hint}) is the Dicke interaction Hamiltonian for superradiance that works via the long-range order of electric dipoles in the system\cite{Dicke}.

The characteristic length for the long-range order of water molecules, that is, the {\it coherence length} (i.e., the wavelength of a resonant photon), denoted by $l_c$, is estimated to be the inverse of the wavenumber of a resonant photon ${\cal E}/(\hbar c)$.
As ${\cal E}/(\hbar c)\approx 160$ [cm$^{-1}$], $l_c\approx 63$ [$\mu$m]\cite{Franks}.
In our model, we assume that the $x$-$y$ dimensions of the system fall within this length $l_c$.

In the semi-classical treatment, the generic form of the Hamiltonian of the water molecule system relevant to our mechanism consists of the rotational kinetic energy of water molecules with average moment of inertia $I^{(w)}=2m_pd_g^2$ (where $m_p$ is the proton mass and $d_g\approx 0.82$ [\AA])\cite{Franks}, the water solvation potential of ions and the interaction Hamiltonian of water molecules with the radiation field and the static electric field:
\begin{widetext}
\begin{equation}
\widehat{H}^{(w)}=\sum_{I=1}^{N}\sum_{i=1}^{n_I}\frac{1}{2I^{(w)}}\widehat{L}_{I,i}^2+\sum_{I=1}^{N}\sum_{i=1}^{n_I}{v}_{I,i}^{(i-w)}+\sum_{I=1}^{N}\sum_{i=1}^{n_I}{v}_{I,i}^{(LJ)}-\sum_{I=1}^{N}\sum_{i=1}^{n_I}\Bigl\{\vec{{A}}\cdot {\vec{\widehat{j}}}_{{I,i}}+\vec{E}_0\cdot \vec{d}_{I,i}\Bigr\}\;,\label{eq:Hw}
\end{equation}
\end{widetext}
where $\vec{d}_{I,i}=(\vec{e}_3)_{I,i}d_0$ and $v_{I,i}^{(i-w)}$ is the screened Coulomb potential between the $I$-th ion and the charges on a water molecule.
The Lennard-Jones potential $v_{I,i}^{(LJ)}=4\epsilon_{LJ}[(\sigma_{LJ}/r_{I,i})^{12}-(\sigma_{LJ}/r_{I,i})^6]$ causes the $I$-th ion be solvated by $n_I$ water molecules.
It is a function of the distance $r_{I,i}$ between the $I$-th ion and the $i$-th water molecule and is attractive for $r_{I,i}>\sigma_{LJ}$ and repulsive for $\sigma_{LJ}>r_{I,i}$, where $\sigma_{LJ}$ is a very short distance of the order of $1$ [\AA]\cite{Water}.
When there is no external electromagnetic field, ${v}_{I,i}^{(i-w)}$ determines the configuration of electric dipoles of water molecules that are in contact with an ion.
The solvation potentials are translationally invariant with respect to phase coordinates $\theta_{I,i}$.
It is a significant point that, in our system, the bulk water molecules that screen the charges of ions do not form part of the laser mechanism, due to thermal noise that prevents ordered motion.

Now, due to the FEL-like mechanism, the collective and coherent behavior of the $x$-$y$ phase coordinates of water molecules follows.
This mechanism consists of two interlocked parts in a positive feedback cycle.
The first part is the long-range collective ordering of dipole vectors of water molecules that solvate each ion moving along the $z$-axis.
As a consequence, the radiation from the clusters of ordered rotating water molecules is almost monochromatic and the time-dependent process of the radiation field and order of water molecules approximates a coherent wave amplified along the $z$-axis.
These approximations are improved by positive feedback in the second part of the mechanism: that is, the FEL-like wave-particle interaction process where an exponential instability of the fluctuation around the dynamic equilibrium state (i.e., our ready state) accompanies both the magnification of the radiation intensity and the water molecule's $XY$-phase bunching that produces long-range ordering\cite{Bonifacio,Kim}.
Finally, we will find that this positive feedback cycle leads to laser radiation.

Under the assumption of monochromaticity of the radiation, the details of the first part of the mechanism are as follows.
A part of the classical radiation field is a transverse wave in the $x$-$y$ plane, which can be written as 
\begin{equation}
A_x+iA_y=A_0e^{-i\phi_0}\;,
\end{equation}
where $A_0$ is positive and real.
The photon-water molecule part of $\widehat{H}_{{\rm int}}^{(p-i-w)}$ can be written as
\begin{eqnarray}
\widehat{H}_{{\rm int}}^{(p-w)}=-\sum_{I=1}^{N}\sum_{i=1}^{n}\vec{{A}}\cdot(\vec{e}_1 \dot{\widehat{d}}_1+\vec{e}_2 \dot{\widehat{d}}_2)_{I,i}\label{eq:Hint0}
\end{eqnarray}
in the representation in which the water molecule's free Hamiltonian is diagonalized by ${\cal E}\widehat{s}_{I,i}^3$.
To approximate this equation, we drop the $I$-dependence of the number $n_I$ ($I=1,2,\ldots,N$) and assume that the permanent polarization of the solvent water-molecule's electric dipoles in the $z$-direction, $d_0^{({{\rm ave}})}$, is uniform over all solvated-ions.
Due to this approximation, the expectation value of Eq.(\ref{eq:Hint0}) can be written using the density matrix $\widehat{\varrho}^{(w)}$ of the system of water molecules as
\begin{eqnarray}
{\rm tr}\Bigl[\widehat{H}_{{\rm int}}^{(p-w)}\widehat{\varrho}^{(w)}\Bigr]\approx \sum_{I=1}^{N}A_0\omega_c{\Delta n}\tilde{d}_0^{({\rm ave})}\frac{1}{2}\sin(\theta_I+\phi)\;,\label{eq:ave}
\end{eqnarray}
where we have introduced new phase variables $\theta_I$ ($I=1,2,\ldots,N$), the shifted phase $\phi=\phi_0+\delta$ with $\delta=\omega_ct$ for resonance angular frequency $\omega_c=c/l_c$, the rescaled permanent polarization $\tilde{d}_0^{({\rm ave})}=d_0^{({\rm ave})} \tilde{d}_0/d_0$, and $\Delta n=n_--n_+$ with $n_+$ and $n_-$ referring to the numbers of excited and ground state water molecules, respectively, arithmetically averaged over all ions.
Here, we assume thermal equilibrium (i.e., the Boltzmann distribution).
Then, we obtain
\begin{eqnarray}
\Delta n=n\tanh\biggl(\frac{{\cal E}}{k_BT}\biggr)\;,
\end{eqnarray}
where ${\cal E}/k_BT\approx 0.12$ at room temperature ($T=300$ [K]) and $\Delta n\approx 3.6$ for $n=30$.

In the following, we will derive Eq.(\ref{eq:ave}).
The basis of the restricted Hilbert subspace of the water molecule's quantum pure states is the set of the symmetrized collective energy spin states\cite{PRSR} with respect to each solvated ion:
\begin{eqnarray}
|L,M\ra&=&\sqrt{\frac{(L+M)!}{n!(L-M)!}}\nonumber\\
&&\times\Biggl\{\sum_{j=1}^n\widehat{s}_j^{(-)}\Biggr\}^{(L-M)}|(e,e,e\ldots e)_n\ra
\end{eqnarray}
with $\widehat{s}_j^{(-)}=\widehat{s}_j^1-i\widehat{s}_j^2$.
In the collective energy spin state $|L,M\ra$, $L$ is an integer or half-integer $n/2$ and $M$ is an integer or half-integer $\Delta n/2$ which runs over $-L\le M\le L$.
Here, to symmetrize the quantum pure state basis of the system of water molecules, which defines the coupling between water molecules and the radiation field, we have used the fact that a photon wavelength of the order of $l_c$ is much longer than the dimensions of the system of one ion and its solvent water molecules, that is, of the order of $1$ [\AA]\cite{Dim}.\cite{PRSR}
With respect to the $j$-th water molecule, the most general forms of the internal parts of the wave functions $\psi^{(g)}_j$ and $\psi^{(e)}_j$ excited and de-excited, respectively, from thermal equilibrium into a set of superradiant states with $M$ having been reduced to $0$ by classical radiation at the $I$-th ion are
\begin{eqnarray}
\psi^{(g)}_j&=&\frac{1}{\sqrt{2}}\Bigl\{|e\ra_je^{i\vartheta_1}-|g\ra_je^{i\vartheta_2}\Bigr\}\;,\\
\psi^{(e)}_j&=&\frac{1}{\sqrt{2}}\Bigl\{|g\ra_je^{-i\vartheta_1}+|e\ra_je^{-i\vartheta_2}\Bigr\}\;,
\end{eqnarray}
where
\begin{equation}
\vartheta_2-\vartheta_1=\theta_I+\delta\;.
\end{equation}
Then we have
\begin{eqnarray}
\sum_{i=1}^{n}\langle (\vec{e}_{1}\dot{\widehat{d}}_{1}+\vec{e}_{2}
\dot{\widehat{d}}_{2})_{I,i}\rangle
&\approx&\omega_{c}{\Delta n}\tilde{d}
_{0}\frac{1}{2}\Biggl[-\frac{1}{n}\Biggl\{\sum_{i=1}^n\vec{e}_{1,i}\Biggr\}\sin (\theta +\delta )
\nonumber
\\
&&{}+\frac{1}{n}\Biggl\{\sum_{i=1}^n\vec{e}_{2,i}\Biggr\}\cos (\theta +\delta )\Biggr]_{I}\;,
\label{eq:dipole}
\end{eqnarray}
where we have used
\begin{eqnarray}
&&\frac{1}{n}\sum_{i=1}^n\vec{e}_{k,I,i}\approx\frac{1}{n_+}\sum_{\bigl\{\psi^{(e)}\bigr\}_I}\vec{e}_{k,I,i}\approx\frac{1}{n_-}\sum_{\bigl\{\psi^{(g)}\bigr\}_I}\vec{e}_{k,I,i}\;,\nonumber\\
&&k=1,2\;.
\end{eqnarray}
Eq.{(\ref{eq:dipole})} leads to Eq.{(\ref{eq:ave})}.
As a significant point here, Eq.(\ref{eq:ave}) is due to purely quantum mechanical processes and arises from the off-diagonal elements of the density matrix (i.e., quantum coherence).

Now, after straightforward calculations, we find the Schr${\ddot{{\rm o}}}$dinger equations for superradiant $\psi^{(g)}_{j}$
($\vec{e}_{k,I,j}$, for $k=1,2$, is reduced to its arithmetic average over $j$ due to our restriction of the Hilbert space) in the interaction picture:
\begin{eqnarray}
\frac{i}{2}\omega_c+i\dot{\vartheta}_{1,I}&=&-\frac{1}{2\hbar}A_0\omega_c\tilde{d}_0^{({\rm ave})}e^{i(\theta_I+\phi)}\;,\\
\frac{i}{2}\omega_c-i\dot{\vartheta}_{2,I}&=&-\frac{1}{2\hbar}A_0\omega_c\tilde{d}_0^{({\rm ave})}e^{-i(\theta_I+\phi)}
\end{eqnarray}
under the two-level approximation with a fixed rotational energy spectrum.
These equations can be combined into
\begin{equation}
i\hbar\dot{\theta}_I=A_0\omega_c\tilde{d}_0^{({\rm ave})}\cos (\theta_I+\phi)\;.
\end{equation}
This equation allows the following pulse form radiation solutions only.
\begin{eqnarray}
\theta_I&=&\theta_0\;,\\
-\phi_0&=&\omega_ct+\theta_0+\frac{\pi}{2}+n\pi\;,
\end{eqnarray}
where $\theta_0$ is time-independent and $n$ is an integer.

As can be seen from this result, to treat the general radiation solutions, we need to relax the rotational energy spectrum of the water molecules in the two-level approximation.
To do this, by a classical mechanical procedure, we change the rotational energy gap ${\cal E}$ as
\begin{equation}
\frac{I^{(w)}\omega_c^2}{2}\to\frac{I^{(w)}(\omega_c+\dot{\theta}_I)^2}{2}
\end{equation}
in $\la\widehat{H}^{(w)}\ra$ such that $\dot{\theta}_I\ll \omega_c$.
According to this change, we can write down the equations of motion of the energy spin system as the canonical equations of $\la\widehat{H}^{(w)}\ra$ with respect to the variables $\theta_I$ and their canonical conjugates $L_I$ and start to describe the second part of the FEL-like mechanism.
(Note that the water molecule's ground state energy is $-{\cal E}/2$.)
Here, we take the origin of the coordinate system to be near the center of the system of radiators.
Then, under the suppression of the second time derivatives of the complex amplitude of the ansatz\footnote{This suppression is attributed to the assumption that the characteristic time for the change of the complex amplitude of $A_{x,y}$ is much longer than the radiation wave period which is of the order of $2\pi/\omega_c\approx 1.3\cdot 10^{-12}$ [s].}
\begin{equation}
(A_x+iA_y)(r,t)={A}(t)\frac{e^{i(ct-r)/l_c}}{r}\;,
\end{equation}
and the suppressions according to the assumptions
\begin{equation}
\dot{\theta}_I\;,\ \ \frac{\dot{A}}{A}\ll \omega_c\;,\label{eq:ll}
\end{equation}
the canonical equations of the phase coordinates and the angular momenta of water molecules and the equation of motion of the radiation field are
\begin{eqnarray}
\frac{n}{2}\bigl(\omega_c+\dot{\theta}_{I}\bigr)&=&\frac{L_{I}}{I^{(w)}}\;,\label{eq:th}\\
\dot{{L}}_{I}&=&-A_0{\omega_c}{\Delta n}\tilde{d}_0^{({\rm ave})}\frac{1}{2}\cos(\theta_I+\phi)\;,\\
-\frac{i\omega_c}{r}\dot{\tilde{{A}}}e^{i\delta}&=&-\sum_{I=1}^{N}\sum_{i=1}^n\mu c^2\tilde{j}_{{I,i}}\label{eq:EEq}
\end{eqnarray}
with $\tilde{{A}}=rA_0e^{-i\phi}$, a complexification of the electric charge current density $\tilde{j}=j_x+ij_y$, and $\mu\approx \mu_0$ being the magnetic permeability in water.
The equation of motion of $A_{x,y}$, Eq.(\ref{eq:EEq}), is equivalent to
\begin{eqnarray}
\dot{A}_0&=&\frac{\mu c^2 N{\Delta n}\tilde{d}_0^{({\rm ave})}}{V}\frac{1}{2}\la\cos (\theta_{I}+\phi)\ra_{I}\;,\label{eq:CE3}\\
\dot{\phi}&=&-\frac{\mu c^2 N\Delta n \tilde{d}_0^{({\rm ave})}}{A_0V}\frac{1}{2}\la \sin (\theta_{I}+\phi)\ra_{I}\;,\label{eq:CE4}
\end{eqnarray}
where $V$ is the volume of the system.
In order to facilitate analysis, we introduce dimensionless variables:
\begin{equation}
{\cal A}_0=A_0\biggl(\frac{\alpha}{2\beta^2}\biggr)^{1/3}\;,\ \ \tau=t\biggl(\frac{\alpha\beta}{2}\biggr)^{1/3}\;,
\end{equation}
where
\begin{eqnarray}
\alpha=\frac{\Delta n\omega_c\tilde{d}_0^{({\rm ave})}}{nI^{(w)}}\;,\ \ \beta=\frac{\mu c^2N\Delta n \tilde{d}_0^{({{\rm ave}})}}{2V}\;.
\end{eqnarray}
We denote the scaled time derivative by a prime.
Then, the set of equations of motion becomes
\begin{eqnarray}
\theta_I^{\prime\prime}&=&-2{\cal A}_0\cos (\theta_I+\phi)\;,\label{eq:EOM1}\\
{\cal A}_0^\prime&=&\la \cos (\theta_I+\phi)\ra_I\;,\label{eq:EOM2}\\
\phi^\prime&=&-\frac{1}{{\cal A}_0}\la \sin(\theta_I+\phi)\ra_I\;,\label{eq:EOM3}
\end{eqnarray}
which is exactly the same as that of the conventional FEL model as described in Ref.\cite{Nature}.

In this paragraph, we qualitatively explain the FEL-like mechanism, following the discussion in Ref.\cite{Nature}.
When the radiation intensity is initially zero, since the phase coordinates $\theta_{I}$ with respect to $I$ are distributed randomly and uniformly within $0\le \theta_I<2\pi$, the system does not change over time.
However, when the initial values of ${\cal A}_0$ and $\phi$ satisfy the conditions $0<{\cal A}_0\ll 1$ and $\phi=0$, respectively, the ponderomotive potential $\Phi=\Phi(\theta)$ for water molecules is subtly perturbed around $\theta=3\pi/2$ due to a small bunching induced by the forces.
At this point, ${\phi}^\prime$ is positive and though $-\la \sin(\theta_{I}+\phi)\ra_{I}\ll 1$ holds, since ${\cal A}_0\ll 1$ holds too, their ratio ${\phi}^\prime$ can be significant.
Then, the phase of the radiation field $\phi$ starts to change and the total phase $\la \theta_{I}\ra_{I}+\phi$ starts to grow from $3\pi/2$.
Then, ${\cal A}_0^\prime$ becomes positive.
Due to this instability, the periodic ponderomotive potential wells $\Phi$ deepen and the bunched water molecules for solvated ions fall to the bottom of $\Phi$ on the space of phase coordinate $0\le \theta<2\pi$.
Then, the rotational kinetic energy of water molecules is transferred to $\Phi$ and this gives a positive feedback cycle: this is the FEL-like mechanism.

Due to this mechanism, when the system reaches ${\cal A}_0\approx 1$, the positive feedback loop is expected to close; the system then enters the non-linear saturated regime.
At the same time, the coherent dynamics of the radiation field and maximally bunched water molecules arises: the quantum coherence of water molecules (i.e., the sum of Eq.(\ref{eq:dipole}) over the system of $N$ ions) is coupled over the system of ion-solvated water, and the intensity of the radiation field is magnified by a multiplicative factor of the order of $N^{4/3}$.
As the conclusive formulae for this regime, we obtain
\begin{eqnarray}
\biggl(\frac{\alpha}{2\beta^2}\biggr)^{-1/3}&=&c_A\cdot \rho^{2/3}\cdot P^{1/3}_z\;,\label{eq:FA}\\
c_A&\approx&2.6\cdot 10^{-22}\ [{\rm m}^3\cdot {\rm kg}\cdot{\rm s}^{-2}\cdot {\rm A}^{-1}]\;,\\
\biggl(\frac{\alpha\beta}{2}\biggr)^{-1/3}&=&c_t\cdot \rho^{-1/3}\cdot P^{-2/3}_z\;,\label{eq:FT}\\
c_t&\approx&2.4\cdot 10^{-4}\ [{\rm m}^{-1}\cdot{\rm s}]\;,
\end{eqnarray}
where $\rho=N/V$ is the ion number concentration in the system and $P_z=d_0^{({\rm ave})}/d_0$ is the permanent electric polarization of water molecules under the static electric field $E_{0,z}$.
The first formula Eq.(\ref{eq:FA}) refers to the value of $A_0$ at the time when ${\cal A}_0\approx 1$, and the second formula Eq.(\ref{eq:FT}) refers to its gain time.
Notably, both formulae do not contain the ion velocity $v$.
Here, according to the formulae in Ref.\cite{GPV}, we obtain the dependence of $P_z$ on $E_{0,z}$ for a realistic value of $E_{0,z}$\footnote{This formula holds for $E_{0,z}\lesssim 10^7$ $[{\rm m}^{-1}\cdot{\rm V}]$ in itself under the setting of Ref.\cite{GPV}.}
\begin{eqnarray}
P_z&\approx&c_P\cdot E_{0,z}\;,\\
c_P&\approx&9.1\cdot 10^{-9}\ [{\rm m}\cdot{\rm V}^{-1}]\;.
\end{eqnarray}

Finally, we have to consider saturation effects.
Our system satisfies the condition 
\begin{equation}
l_b\ll l_s={(c-v)}\frac{l_g}{v}\;,
\end{equation}
where $l_b$ and $l_g$ are the bunch and gain lengths, respectively.
In this case, the radiation emitted by a sufficiently small bunch of water molecules could quickly escape from it due to slippage; so saturation effects would be reduced\cite{FELSR1,FELSR2}.
In a dissipative system whose original system is governed by the same equations of motion (i.e., Eqs.(\ref{eq:EOM1}) to (\ref{eq:EOM3})) as our system, it has been argued that superradiation, avoiding saturation effects, can be realized\cite{FELSR1,FELSR2}.
However, in our system its growth rate is too slow due to the largeness of the ratio of the slippage distance $l_s$ to the bunch length $l_b$; so this superradiance scenario {for the system of all ions} has to be abandoned.

Now, we summarize the overall results.

In this Letter, we studied the model of a resonant system of photons and ion cluster-solvated rotating water molecules for the case where ions in the cluster have very low, non-relativistic velocities and direction of motion parallel to a static electric field induced in a single direction.
The system of water molecules solvating an ion is reduced to an $XY$ energy spin system under the two-level approximation in the rotational spectrum.
By incorporating the mechanism of Dicke superradiation {for each ion}, we found that the equations of motion of the $XY$ energy spin systems, over all ions, coupled to the radiation field are expressed in terms of a conventional free electron laser.
This result leads to a dynamical coherence mechanism, induced by collective instability in the wave-particle interaction.

In the application of our result, a key earlier result is Ref.\cite{GPV}, which shows the existence of a permanent electric polarization $P_z$ of water molecules under a strong static electric field induced in one direction.
This may occur, for example, around an impurity that carries a sizeable electric dipole.
We use this earlier result for the $XY$ spin picture of water molecules.
  In conclusion, our coherence mechanism has desirable properties when it is applied to models of the cluster current of a large number of ions solvated in water.
As described earlier, this is assumed to be under a strong static electric field induced parallel to the ion current, with the dimensions of the ion cluster falling within the coherence length (i.e., the wavelength of a resonant photon) $l_c$, and the condition Eq.(\ref{eq:ll}) must also be satisfied.

\end{document}